\documentclass[10pt,twocolumn]{paper}
\usepackage{epsfig, graphicx}
\usepackage[english]{babel}
\usepackage[margin=1.5cm]{geometry}

\title{\center \rm \bf X-Ray Scattering near a  Liquid - Vapor
Phase Transition at the n-Hexane - Water Interface}

\author{\small \rm Aleksey M. Tikhonov\/\thanks{tikhonov@kapitza.ras.ru}\\ \small
Kapitza Institute for Physical Problems, Russian Academy of Sciences, Moscow, Russia}

\begin{document}
\maketitle
%\centerline{\today}

\abstract{ \it The molecular structure of neutral n-triacontanol mesophases at the n-hexane - water interface has been studied by diffuse X-ray scattering using synchrotron radiation. According to the experimental data, a transition to the multilayer adsorption of alkanol occurs at a temperature below the transition from a gas phase to a liquid Gibbs monolayer.}

\vspace{0.25in}
%\large

An amphiphilic adsorption film at an oil -- water interface can be considered as a two-dimensional
thermodynamic system where phase transitions between surface mesophases can occur [1–5]. One of
such systems with the parameters ($p,\,T,\,c$), where $p$ is the pressure, $T$ is the temperature, and $c$ is the concentration of the surfactant in the bulk, is a solvable film of high-molecular-weight n-alkanol at the n-alkane -- water interface [6–8]. A two-dimensional liquid -- vapor thermotropic phase transition is
observed in it at $p=1$\,atm and $c=const$ [9, 10]. In this work, we study the temperature dependence of the
intensity of diffuse (nonspecular) scattering of 15-keV photons at the n-hexane - water interface with an n-triacontanol (C$_{30}$-alcohol) film where such a transition occurs [11]. The analysis of scattering data within the theory of capillary waves shows that, with decreasing temperature, a transition to multilayer adsorption occurs at $T^*$ below the temperature $T_c$ of a two-dimensional transition of condensation of triacontanol to a liquid monolayer. We propose a model of the structure of the interface that makes it possible to relate new diffuse scattering data to previously reported and supplemented reflectometry data.

In this work, we study systems with the bulk concentration $c\approx0.7$\,mmol/kg of C$_{30}$-alcohol in n-hexane. Samples with a $75$\,mm$\times$\,150\,mm  saturated hydrocarbon -- water macroscopically flat interface were prepared and studied in a hermetic stainless-steel cell with polyester windows transparent to X-rays according to the technique described in [12, 13]. The temperature of the cell was controlled by an original two-stage thermostat.

C$_{30}$-alcohol or C$_{30}$H$_{62}$O was purified twice by recrystallization at room temperature from a supersaturated solution in n-hexane, which was prepared by solving n-triacontanol in n-hexane at a temperature of $T\approx 333$\,K [14]. Saturated hydrocarbon C$_6$H$_{14}$ (the melting temperature is $\approx 178$\,K, the boiling temperature is $\approx 342$\,K, and the density at 298\,K is $\approx 0.65$\,g/cm$^3$) was preliminarily purified by repetitive filtering in a chromatographic column [15]. Deionized water (Barnstead, NanoPureUV) was used as the bottom bulk phase, in which C$_{30}$H$_{62}$O is almost insoluble. To prevent the formation of gas bubbles at the interface, the sample was "annealed": the temperature of liquids in the cell was increased up to $T \cong 330$\,K and then was reduced below $T_c$.

According to [11, 16] the liquid -- vapor transition in the C$_{30}$H$_{62}$O neutral absorption film at the n-hexane - water interface at the pressure $p=1$\,atm and concentration $p=1$\,àòì è $c\approx0.7$\,mmol/kg occurs at $T_{c}\approx 301$\,K. The corresponding temperature dependence of the interfacial tension $\gamma(T)$ for this system is shown by circles in Fig. 1. The change in the slope of $\gamma(T)$ is related to the change in the surface enthalpy upon the transition: $\Delta H = - T_c\Delta(\partial \gamma/\partial T)_{p,c}$ $=1.3\pm 0.1$\,J/m$^2$. With increasing temperature near $T_{c}$, a large number of adsorbed C$_{30}$-alcohol molecules leave the interface and are dissolved in the bulk of the hydrocarbon solvent. In this process, the surface concentration of alcohol molecules decreases by more than an order of magnitude and a gas amphiphilic mesophase is achieved.

\begin{figure}
\hspace{0.1in}
\epsfig{file=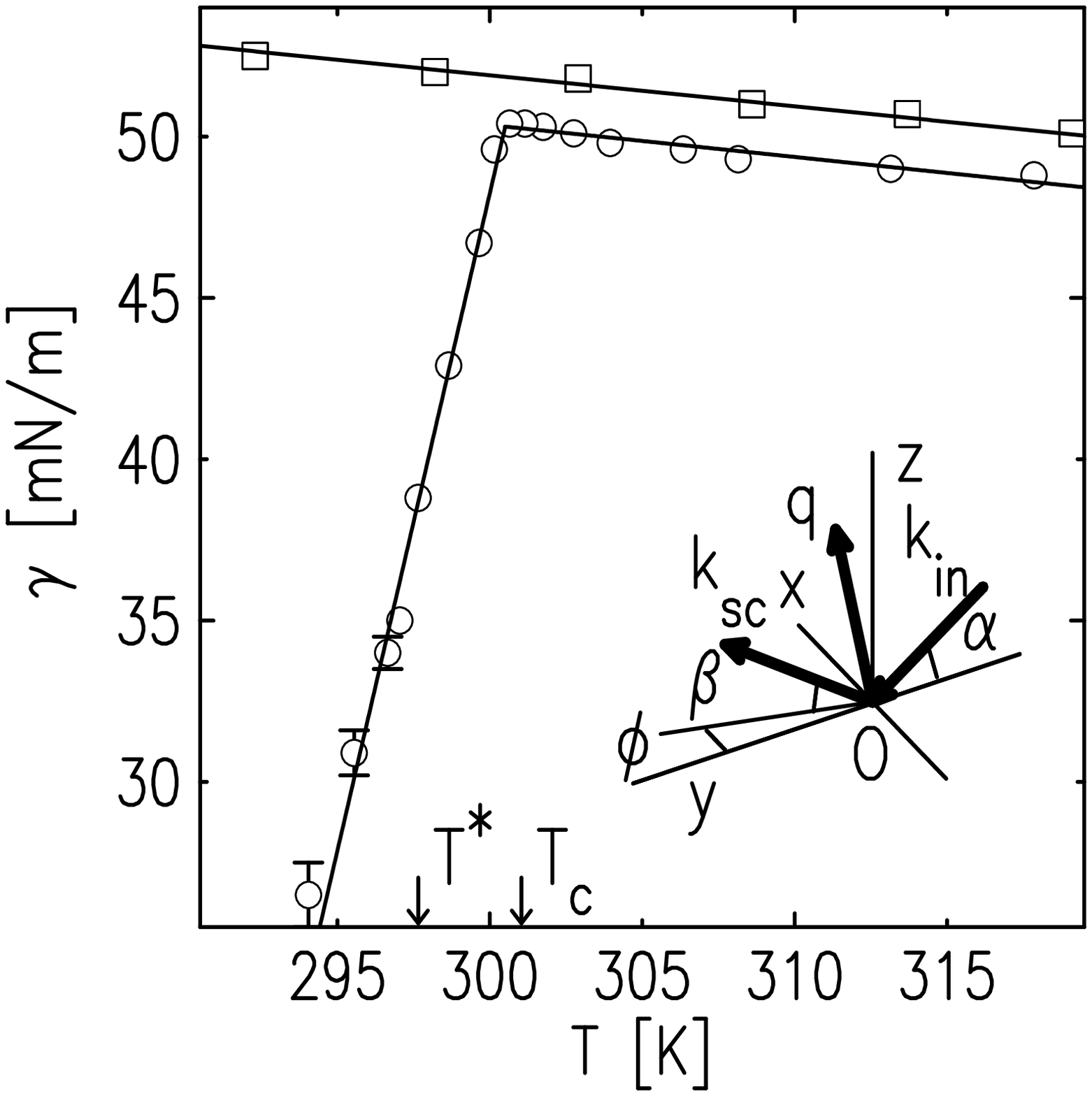, width=0.35\textwidth}

\small {\bf Figure 1}  Temperature dependences of the interfacial tension $\gamma(T)$ at (squares) the pure n-hexane - water interface and (circles) this interface with an adsorbed n-triacontanol layer. A sharp change in the slope of $\gamma(T)$ is observed at the liquid -- vapor phase transition in the C$_{30}$H$_{62}$O adsorption film at $T_{c}\approx 301$\,K. A transition to multilayer adsorption occurs at $T^*\approx 298$\,K. The inset shows the scattering kinematics at the n-hexane - water interface. The plane $xy$ coincides with the interface, the axis $Oõ$ is perpendicular to the beam direction, and the axis $Oz$ is perpendicular to the surface and is directed opposite to the gravitational force.

\end{figure}

The qualitative model of the structure of interfaces shown in Fig. 2 makes it possible to match the previous
reflectometry data to new scattering data for this system. This model is consistent with the structure of a
C$_{30}$H$_{62}$O linear chain molecule with a length of $\approx 41$\,\AA{}($\approx 1.5$\,\AA (-CH$_3$) + $29\times 1.27$\,\AA{}(Ñ-Ñ) + 2.4\,\AA{}(-CH$_2$OH)). As was shown in [11], reflectometry data for the condensed triacontanol mesophase (a Gibbs liquid monolayer) can be explained by a structure consisting of three layers. Polar head parts -CH$_2$OH are involved in the formation of layer 1, whereas layers 2 and 3 are formed by hydrophobic hydrocarbon tails -C$_{29}$H$_{59}$9. Further, it was shown that additional thick layer 4 is necessary for explaining an increase in the intensity of diffuse scattering by the absorption film with decreasing temperature.

The reflection coefficient $R$ and the intensity of diffuse surface scattering of X-rays $I_n$ at the n-hexane - water interface were measured with the use of synchrotron radiation at the X19C beamline of the National Synchrotron Light Source (NSLS, Brookhaven National Laboratory, United States) [17]. The intensity
$I_0$ of the incident monochromatic photon beam at a wavelength of $\lambda=0.825 \pm 0.002$\,\AA{} was $\sim 10^{10}$\, photon/s. This method was recently used to study thermotropic transitions in protonated and ionized melissic acid absorption films at the n-hexane - water interface [18, 19].

Let {\bf k}$_{\rm in}$ and {\bf k}$_{\rm sc}$ be the wave vectors with amplitude $k_0= 2\pi/\lambda$ of the incident and scattered beams, respectively. In a reference frame where the origin $O$ is at the center of the irradiated region, the plane $xy$ coincides with the water boundary, the axis $Ox$ is perpendicular to the beam direction, and the axis $Oz$ is normal to the surface and is directed opposite to the gravitational force. In the experiment, the glancing angle in the plane is $\alpha << 1$, the scattering angle is $\beta << 1$, and the angle in the $xy$ azimuthal plane between the direction of the incident beam and the direction of scattering is $\phi \approx 0$ (see the inset of Fig. 1). In this reference frame, the components of the scattering vector {\bf q = k$_{\rm in}$ {\rm -} k$_{\rm sc}$} in the interface plane $xy$ are $q_x \approx k_0\phi$ and $q_y\approx k_0(\alpha^2-\beta^2)/2$ and the normal component along the axis $Oz$ is $q_z\approx k_0(\alpha+\beta)$.

\begin{figure}
\hspace{0.3in}
\epsfig{file=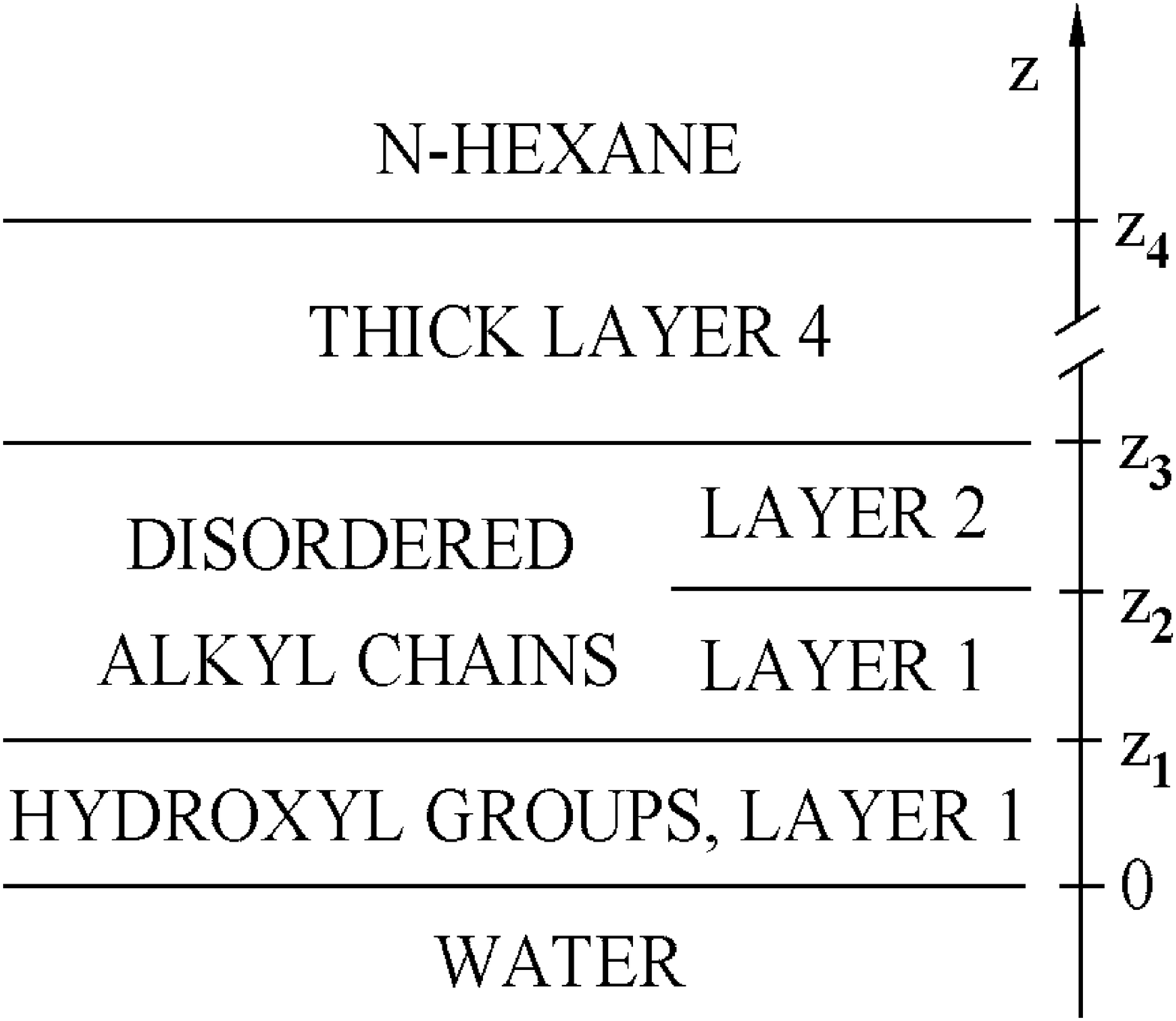, width=0.40\textwidth}

\small {\bf Figure 2} Model of the n-triacontanol C$_{30}$H$_{62}$O adsorption film at the n-hexane - water interface.

\end{figure}

The total external reflection angle $\alpha_c=\lambda\sqrt{r_e\Delta\rho/\pi}$ $\approx 8\cdot10^{-4}$\,rad (where $r_e =2.814\cdot10^{-5}$\,\AA{} is the classical electron radius) for the n-hexane - water interface is determined by the difference $\Delta\rho=\rho_w-\rho_{h}$ between the bulk electron densities in hydrocarbon solvent $\rho_{h}$$\approx0.22$\,{\it e$^-$/}{\AA}$^3$ and in water $\rho_w\approx0.33$\,{\it e$^-$/}{\AA}$^3$.

Figure 3 shows the dependence $R(q_z)$ at various temperatures above and below the phase transition at
the n-hexane - water interface. At $q_z < q_c=(4\pi/\lambda)\alpha_c$$\approx 0.012$\,\AA$^{-1}$, the incident beam undergoes total external reflection; i.e., $R\approx 1$. The inset of Fig. 3 shows the temperature dependence of the normalized reflection coefficient $R/R_1$ measured near the interference maximum at (circles) $q_z=0.2$  and (squares) $q_z=0.225$\,\AA$^{-1}$, where $R_1$ is the reflection coefficient at $T \approx 300.0$\,K. The data presented in Fig. 3 demonstrate that the character of the reflection curves changes sharply with increasing temperature in the range $\Delta T \sim 2$\,K near the temperature $T_{c}$.

Figure 4 shows the data for the normalized intensity of surface scattering $I_n (\beta) \equiv (I(\beta)-I_b(\beta))/I_0$ (normalization condition $I_n(\alpha)\equiv 1$) measured in the temperature range from 295 to 302\,K. Here, $I(\beta)$ is the number of photons scattered by the sample and reflected (specularly and diffusely) from the surface in the irradiated region with an area of $A_0\approx 30$\,mm$^2$ at the center of the interface in the direction $\beta$; $I_0$ is the normalization constant proportional to the intensity of the incident beam, which was controlled in the experiment immediately before the entrance of the beam into the cell; and $I_b(\beta)$ is the number of photons scattered in the bulk of n-hexane on their way to the interface. The method of $I_b$ determining is described in detail in [18]. The most intense peak on the curves corresponds to specular reflection at $\beta = \alpha$, and the peak against the diffuse background at $\beta \to 0$ illustrates an increase in the intensity of scattering at$\beta = \alpha_c$ ($\approx 0.05^\circ$) [19, 20].

\begin{figure}
\hspace{0.3in}
\epsfig{file=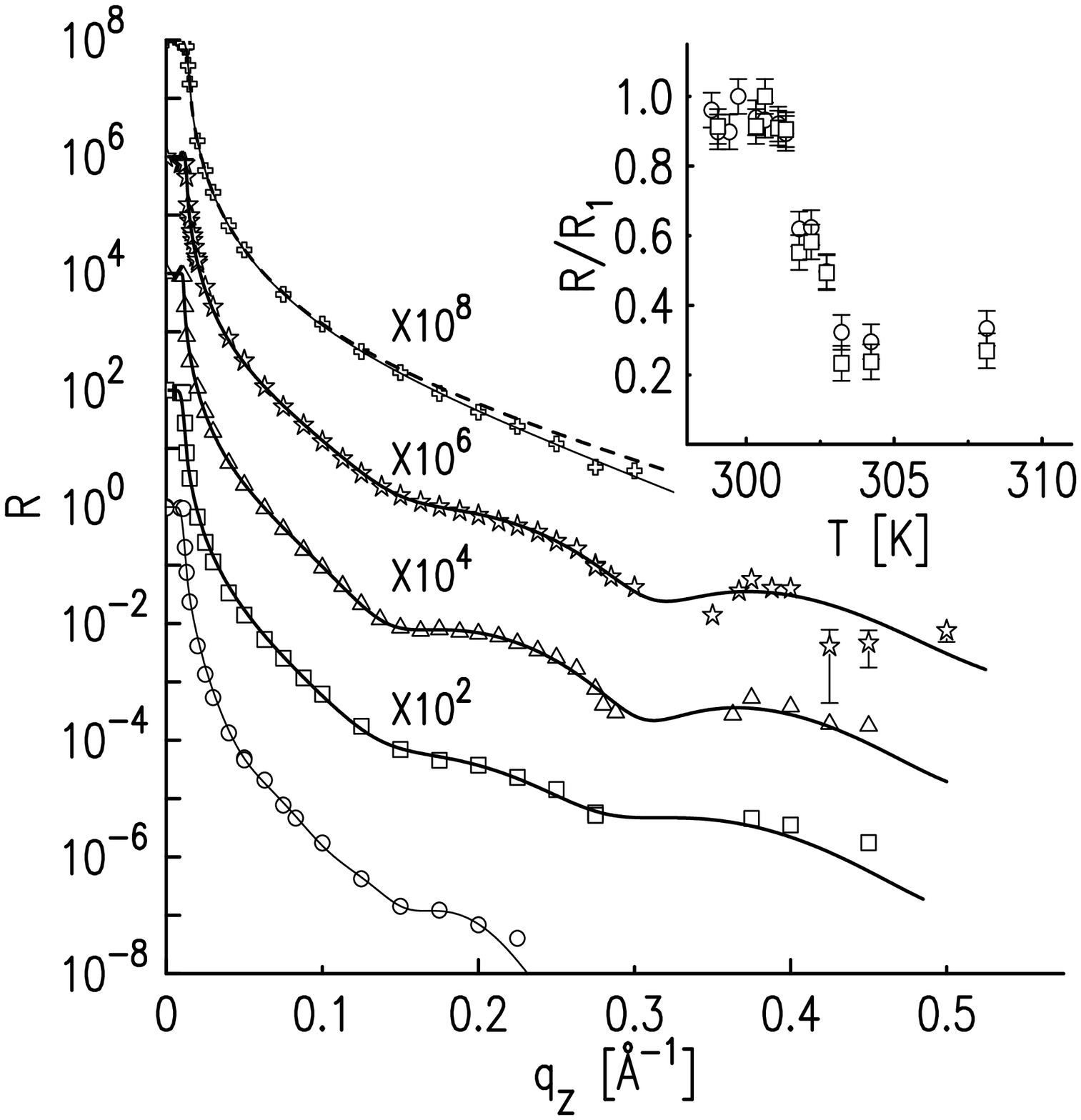, width=0.40\textwidth}

\small {\bf Figure 3} Reflection coefficient $R$ versus $q_z$ for the n-hexane - water interface at temperatures of (circles) 294.8, (squares) 297.2, (triangles) 297.3, (stars) 297.7, and (crosses) 318.2\,K. The lines correspond to models of the adsorbed layer. The inset shows the temperature dependence of the normalized reflection coefficient $R/R_1$ at $q_z=0.2$ (circles) and $q_z=0.225$\,\AA$^{-1}$ (squares), where $R_1$ is the reflection coefficient at $T \approx 300.0$\,K.

\end{figure}

\begin{figure}
\hspace{0.3in}
\epsfig{file=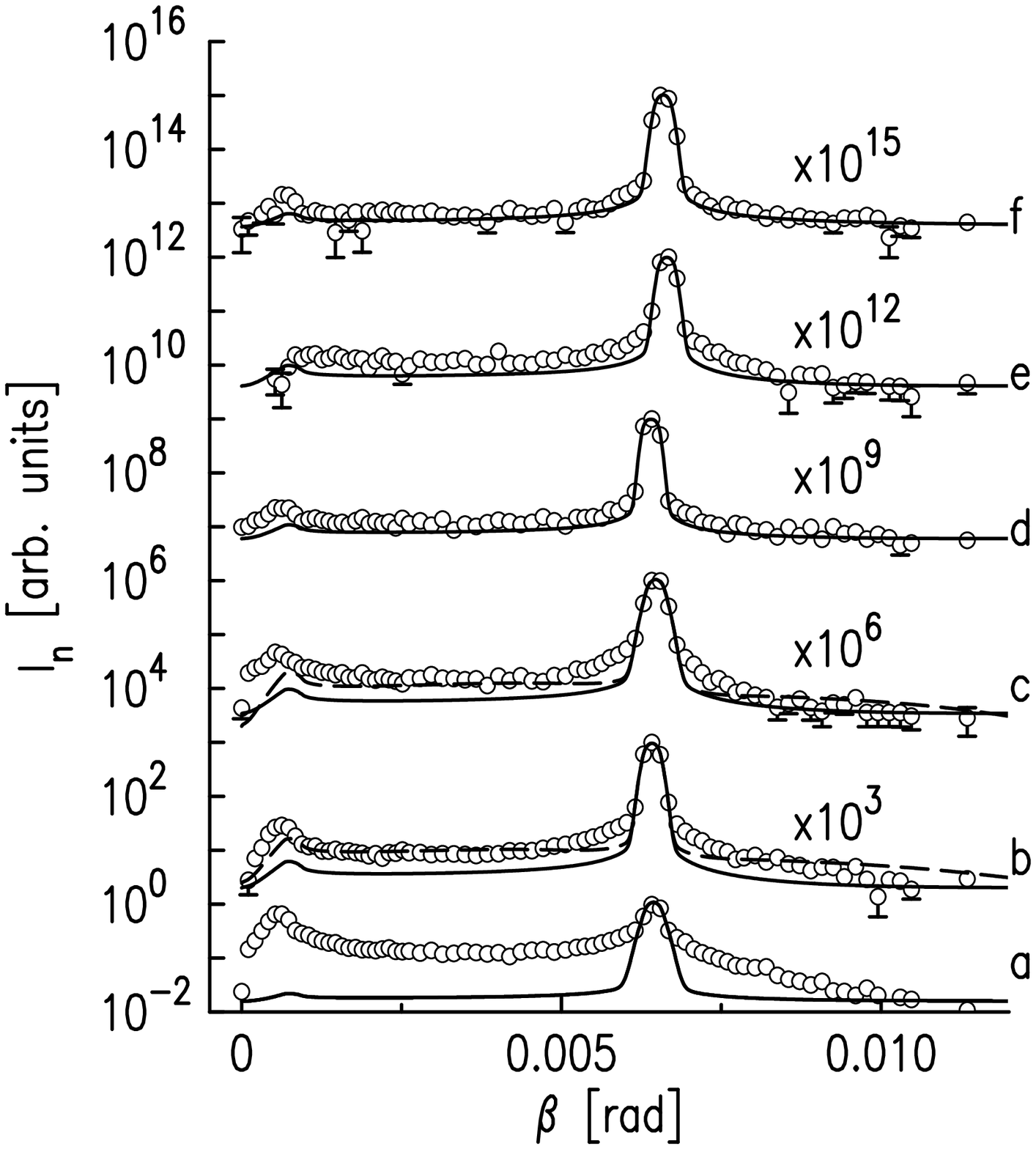, width=0.40\textwidth}

\small {\bf Figure 4} Surface intensity of scattering $I_n$ at a glancing angle of $\alpha \approx 6.6 \cdot 10^{-3}$ at the n-hexane - water interface at temperatures T = (a) 294.8, (b) 297.2, (c) 297.7, (d) 300.3, (e) 301.3, and (f) 301.8\,K. Solid lines show the results of calculations within the monolayer model given by Eq. (6)). Dashed lines show the results of calculations in the model with an extended layer specified by Eq. (8).

\end{figure}

Line $a$ in Fig. 4 corresponds to the liquid mesophase at a low temperature at which the peak of surface
diffuse scattering at small $\beta$ values reaches the specular reflection peak. As the temperature is increased to $T^*\approx 298$\,K, the intensity $I_{\rm diff}$ of scattering decreases gradually to $\approx 2 \cdot 10^{-3}$ of the reflection peak and is then almost independent of $T$ (lines $d,e,f$).

The electron density $\langle \rho(z) \rangle$ of the sample in the direction perpendicular to its surface averaged over the area $A_0$ can be reconstructed from the reflectometry and diffuse scattering experimental data. In the distorted wave Born approximation, this problem is reduced to the determination of the model structure factor $\Phi(q)$ of the form [21]
\begin{equation}
\Phi(q)=\frac{1}{\Delta\rho}\int^{+\infty}_{-\infty}\left\langle\frac{d\rho(z)}{dz}\right\rangle e^{iqz} dz.
\end{equation}

To explain this experiment, it appears sufficient to use the model of interface in the form of a heterogeneous-layered structure (see Fig. 2). However, in the general case, where, e.g., intraplane inhomogeneities of the surface are smaller than the projection of the spatial coherence length of incident radiation on the $xy$ plane ($\lambda^2/(\Delta\lambda\alpha) \sim 10^{-4} - 10^{-6}$\,m), the scattering data can be interpreted only within complex models of spatially inhomogeneous structures [22-25]. 

The intensity of diffuse surface scattering $I_n(\beta)$ on the thermal fluctuations (capillary waves) at the interface was calculated by the method presented in [18, 19, 26, 27].

The dependences $R(q_z)$ and $I_n(\beta)$ in the gas mesophase of the monolayer at $T > T_c$ are quite well described within the single-parameter model of the pure interface with the structure factor
\begin{equation}
\Phi(q) = e^{-\sigma^2_0 q^2/2}.
\end{equation}

The electron density profile corresponding to the structure factor (2) has the form
\begin{equation}
\displaystyle
\langle \rho(z) \rangle =\frac{1}{2}(\rho_w+\rho_h)+\frac{1}{2}(\rho_w-\rho_h){\rm erf}\left(\frac{z}{\sigma_0\sqrt{2}}\right).
\end{equation}

The minimum value of the parameter $\sigma^2_{0}$, which determines the width of the interface, is limited by the "capillary width" squared:
\begin{equation}
\sigma_{0}^2 =  \frac{k_BT}{2\pi\gamma(T)} \ln\left(\frac{Q_{max}}{Q_{min}}\right),
\end{equation}
which is in turn specified by the short-wavelength limit in the spectrum of capillary waves $Q_{max} = 2\pi/a$ (where $a\approx 10$ {\AA} is the intermolecular distance) and $Q_{min}=q_z^{max}\Delta\beta/2$ ($\Delta\beta$ is the angular resolution of the detector and $q_z^{max} \approx 0.5$ {\AA}$^{-1}$) [27, 30-34].

The parameters $R(q_z)$ and $I_n$ calculated by Eq. (2) at $T>T_c$ are shown in Figs. 3 and 4, respectively, by solid lines, which describe the corresponding experimental data within the statistical error of the measurements (crosses in Fig. 3 and line f in Fig. 4). The corresponding fitting parameter $\sigma_0 = (4.4\pm 0.3)$\,\AA{} is insignificantly larger than the value previously measured for the pure n-hexane - water interface [35]. The dashed line in Fig. 3 was calculated without fitting parameters by Eq. (2) with the value $\sigma_0 = 3.45$\,\AA{} calculated by Eq. (4).

The monomolecular liquid n-triacontanol mesophase at $T < T_c$ is described within a qualitative three-layer model (see Fig. 2) proposed in [11] with the structure factor
\begin{equation}
\Phi(q)_m = \frac{e^{-\sigma_0^2q^2/2}}{\Delta\rho}\sum_{j=0}^{3}{(\rho_{j+1}-\rho_j) e^{-iq_zz_j}},
\end{equation}
where $z_0=0$, $\rho_0=\rho_w$, and $\rho_4 = \rho_h$. In the liquid phase, the electron densities are $\rho_1 \approx 1.13\rho_w$, $\rho_2 \approx 0.95\rho_w$, and $\rho_3 \approx 0.79\rho_w$ and the coordinates of the layer boundaries are $z_1 \approx 5$\,\AA{}, $z_2 \approx 18$\,\AA{}, and $z_3 \approx 36$\,\AA. The total thickness of the monolayer is $(36 \pm 2)$\,\AA.

The model electron density profile corresponding to the structure factor (5) has the form
\begin{equation}
\displaystyle
\langle \rho(z) \rangle_m =\frac{1}{2}(\rho_0+\rho_4)+\frac{1}{2}\sum_{j=0}^3(\rho_{j+1}-\rho_j){\rm erf}\left(\frac{l_j}{\sigma_0\sqrt{2}}\right),
\end{equation}
where $l_j=z+z_j$.

The calculated values $\sigma_0$ range from 3\,\AA\,to 6\,\AA{} and coincide with the fitting values within errors.

The parameters $R(q_z)$ and $I_n$ calculated by Eq. (5) are shown in Figs. 3 and 4, respectively, by solid lines, which describe quite well the experimental data near $T_c$ (lines $d$ and $e$ in Fig. 4). However, the observed intensity of scattering at $T < T^*\approx 298$\,K is more than five times larger than the calculated value.

This increase in the intensity of scattering can be qualitatively explained within the four-layer model
(multilayer absorption) and the structure factor [18]
\begin{equation}
\displaystyle
\Phi(q)^*_m +\frac{ \displaystyle \delta\rho e^{-\sigma^2q_z^2/2 }}{ \displaystyle \Delta\rho }  e^{-iq_zz_4}.
\end{equation}

The second term describes the fourth adsorbed layer with the thickness $z_4-z_3$ and the density $\rho_h + \delta\rho$. The parameter $\sigma$ is the intrinsic width of the interface between this layer and the bulk of n-hexane, and $\Phi(q)^*_m$ is given by Eq. (5) with the substitution $\rho_4 = \rho_h + \delta \rho$. With an increase $T \to T^*$, the excess surface concentration is $\delta\rho(z_4-z_3)\to 0$. The distribution corresponding to the structure factor (7) has the form
\begin{equation}
\langle \rho(z) \rangle=\langle \rho(z) \rangle_m +\delta\rho {\rm erf}\left(\frac{z+z_4}{\sigma_0\sqrt{2}}\right)
\end{equation}

The analysis of the data for diffuse scattering and $R(q_z)$ with the use of Eq. (7) is shown in Fig. 4 by dashed lines $b$ and $c$. The estimated thickness of the thick layer is $z_4-z_3 \approx 80$\,\AA,
the parameter is $\delta\rho \approx 0.03 \rho_w$, and the width is $\sigma \approx 30$\,\AA. In this case, the parameters of the liquid monolayer are insignificantly corrected within the error because the contribution of the second term in Eq. (7) drops rapidly with $q_z$ increasing and becomes negligible at $q_z> 0.075$\,\AA$^{-1}$. The electron density in the fourth layer $\rho_h + \delta\rho \approx 0.72 \rho_w$ corresponds to the density in a high-molecular-weight alkane liquid [36].

The low-temperature limit of measurement of surface scattering is bounded by the region $\gamma > 25$\,mN/m.
At lower $\gamma$ values, n-triacontanol aggregates with linear dimensions $< 1$\,mm are precipitated at the n-hexane - water interface. It is reasonable to attribute this behavior to the supersaturation of the solution of C$_{30}$-alcohol in n-hexane at a decrease in the temperature, which is also accompanied by the thickening and densification of layer\,4. In this case, it is difficult to correctly separate the intensities of surface and bulk scattering. For this reason, the model description of the data obtained at $T=294.8$\,K in Fig. 3 (circles) and Fig. 4 (line $a$) is an ill-posed problem.

Thus, the gas mesophase of the monolayer  is implemented at the interface at $T > T_c$. It is characterized
by the single parameter $\sigma_0= 4.4\pm 0.3$\,\AA{} (structure I) and is slightly different from the structure of the pure n-hexane - water interface [12, 35]. In the interval $T^* <T < T_c$, structure II is described by the three-layer model of the monolayer with a thickness of $(36\pm2)$\,\AA, and structure III of the interface at $T< T^*$ is described by the four-layer model and consists of the n-triacontanol monolayer and the high-molecular-weight alkane liquid layer with a thickness of $\sim 100$\,\AA. The proposed model of the interface structure in the range $T<T^*$ makes it possible to consistently describe the experimental data. Model electron density profiles $\langle \rho(z) \rangle$ for mesophases of the absorbed C$_{30}$-alcohol film normalized to $\rho_w$ are shown in Fig. 5.

It is noteworthy that the intensity of diffuse scattering background at the n-hexane - water interface in the transition region at $T\approx T_c$ (line 5 in Fig. 4) exceeds the calculated $I_n(\beta)$ values by $\sim 100 \%$ for both the homogeneous monolayer and the pure interface. This relation can be apparently attributed to the separation of the surface into low- and high-temperature phase domains. Both phases tend to be mixed because the formation of one-dimensional interfaces significantly reduces the interfacial energy [37]. This formally excludes the existence of a first-order phase transition in this system [38, 39]. The structure of the surface is determined by the competition between the long-range electrostatic repulsion and short-range van der Waals interactions [40]. The optical data indicate the existence of an equilibrium spatially inhomogeneous structure of the n-hexane - water interface near $T_c$ [25, 41, 42]. Small-angle grazing scattering by inhomogeneities of the interface can make an additional contribution to diffuse background in the vicinity $\Delta T$ of the transition temperature. However, its quantitative description requires the model of spatial structure, the development of which is beyond the scope of this work [43].

To summarize, the molecular structure of the neutral adsorbed n-triacontanol layer at the n-hexane - water interface in its different phase state has been studied by diffuse scattering and reflectometry using
synchrotron radiation. The calculation of the intensity of scattering for model structures proposed in [11, 16] without fitting parameters satisfactorily reproduces the experimental data near $T_c$. The main and quite surprising result of the analysis of the data is that the transition to multilayer adsorption occurs at the temperature $T^*$ below the temperature $T_c$ of the liquid -- vapor transition in the absorption film. The observation of such a transition in a protonated melissic acid layer above its melting temperature was previously reported, although the dependences $\gamma(T)$ near $T^*$ for these systems are strongly different [18].

The work at the National Synchrotron Light Source was supported by the US Department of
Energy (contract no. DE-AC02-98CH10886). The work at the X19C beamline was supported by the
ChemMatCARS National Synchrotron Resource, University of Chicago, University of Illinois at Chicago,
and Stony Brook University.

\begin{figure}
\hspace{0.3in}
\epsfig{file=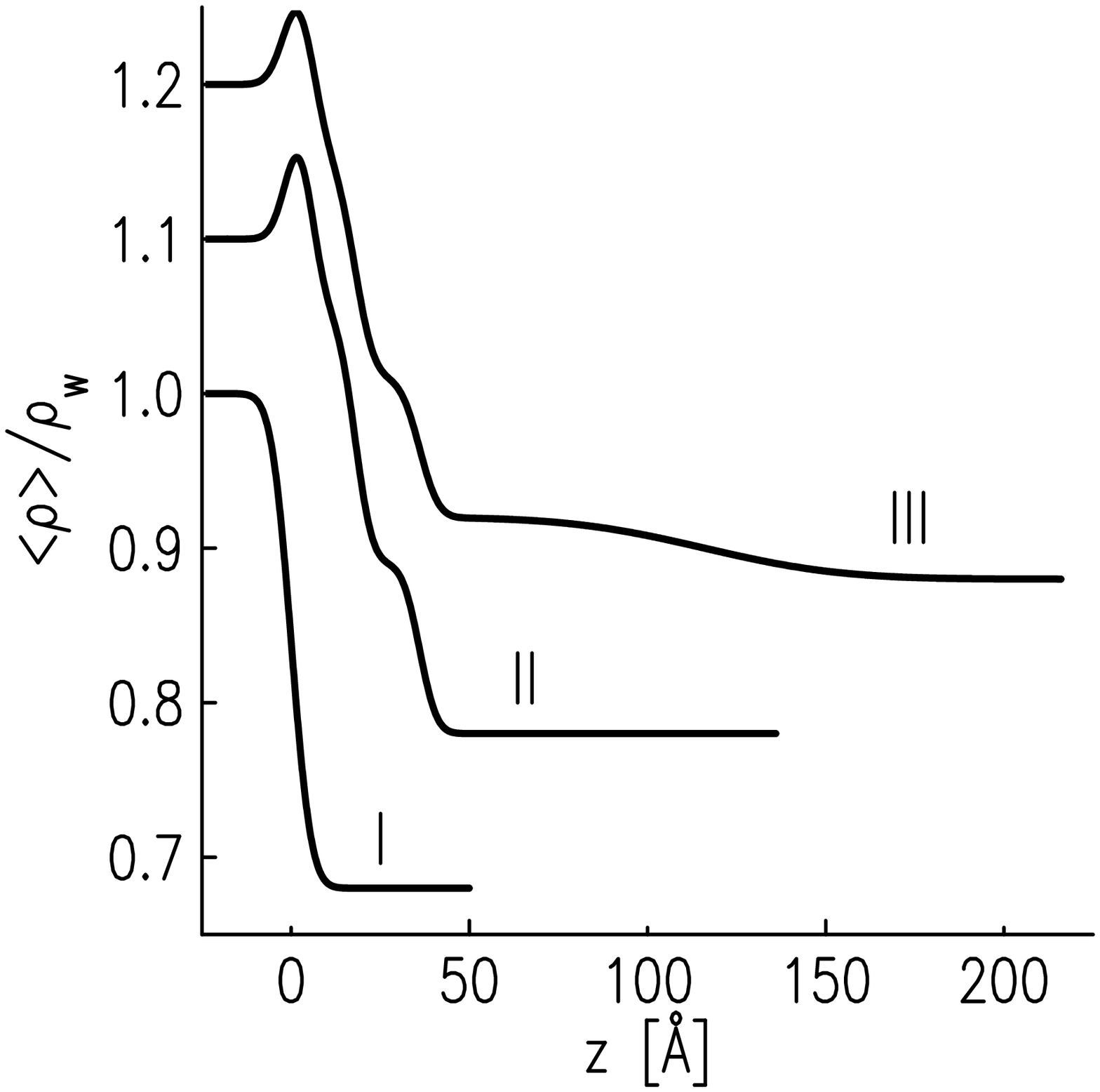, width=0.40\textwidth}

\small {\bf Figure 5} Model profiles of the electron density $\langle \rho(z) \rangle$ of the C$_{30}$-alcohol absorption film normalized to that in water under normal conditions ($\rho_w=0.333$ {\it e$^-$/}{\AA}$^3$): (I) the model of the gas phase at $T>T_c$ and $\sigma_0= 4.4$\,\AA{}, (II) the
three-layer model of the monolayer at $T^*<T<T_c$ and $\sigma_0= 3.8$\,\AA{}, and (III) the four-layer model with a liquid monolayer at $T<T^*$, $\sigma_0=4.2$\,\AA{} and $\sigma=30$\,\AA{}. For convenient comparison, profiles II and III are shifted along the y axis by 0.1 and 0.2, respectively. The position of the interface between the polar region of n-triacontanol molecules and water is placed at $z=0$.
\end{figure}


\begin{thebibliography}{49}
\small

\bibitem{Gibbs}
J. W. Gibbs, {\it Collected Works, Vol. 1}, p. 219. Dover, New York, 1961.

\bibitem{Adamson}
A.\,W.\,Adamson, {\it Physical Chemistry of Surfaces}, 3rd ed.; John Wiley \& Sons: New York, 1976.

\bibitem{LL5}
L.\,D.\,Landau and E.\,M.\,Lifshitz, {\it Course of Theoretical Physics}, Vol. 5: {\it Statistical Physics} (Nauka, Moscow, 1995; Pergamon, Oxford, 1980).

\bibitem{Hansen}
R.\,S.\,Hansen, J. Phys. Chem. 66, 410 (1962).

\bibitem{Motomura1}
K.\,Motomura, J Colloid Interface Sci. 64, 348 (1978).

\bibitem{Motomura}
K.\,Motomura, N.\,Matubayasi, M.\,Aratono, R.\,Matuura, J. Colloid Interface Sci. 64, 356 (1978).

\bibitem{Baret}
M.\,Lin, J.\,L\,Ferpo, P.\,Mansaura, J.\,F.\,Baret, J. Chem. Phys. 71, 2202 (1979).

\bibitem{Aratono}
M.\,Aratono, T.\,Takiue, N.\, Ikeda, A.\,Nakamura, K.\,Motomura, J. Phys. Chem. 97, 5141 (1993).

\bibitem{Takiue1}
T.\,Takiue, T.\,Matsuo, N.\,Ikeda, K.\,Motomura, M.\,Aratono, J. Phys. Chem. B 102, 4906 (1998).

\bibitem{TAMSCH}
A.\,M.\,Tikhonov, M.\,L.\,Schlossman, J. Phys.: Condens. Matter 19, 375101 (2007).

\bibitem{TAMSCH1}
A.\,M.\,Tikhonov, M.\,L.\,Schlossman, J. Phys. Chem. B 107, 3344 (2003).

\bibitem{s-cell}
D.\,M.\,Mitrinovic, Z.\,J.\,Zhang, S.\,M.\,Williams, Z.\,Q.\,Huang and M.\,L.\,Schlossman, J. Phys. Chem. B 103, 1779 (1999).

\bibitem{Tikh311}
A.\,M.\,Tikhonov, J. Chem. Phys 124, 164704 (2006).

\bibitem{Takiue}
T.\,Takiue, A.\,Yanata, N.\, Ikeda, K.\,Motomura, M.\,Aratono, J. Phys. Chem. 100, 13743, (1996).

\bibitem{alkanes}
A.\,Goebel, K.\,Lunkenheimer, Langmuir 13, 369 (1997).

\bibitem{TAMPINSCH}
A.\,M.\,Tikhonov, S.\,V.\,Pingali, M.\,L.\,Schlossman, J. Chem. Phys. 120, 11822 (2004).

\bibitem{x19c}
M.\,L.\,Schlossman, D.\,Synal, Y.\,Guan, M.\,Meron, G.\,Shea-McCarthy, Z.\,Huang, A.\,Acero, S.\,M.\,Williams, S.\,A.\,Rice, P.\,J.\,Viccaro, Rev. Sci. Instrum. 68, 4372 (1997).

\bibitem{acid-c30-3}
A.\,M.\,Tikhonov, JETP Lett. 104, 309 (2016). 

\bibitem{acid-c30-4}
A.\,M.\,Tikhonov, JETP Lett. 105, 775 (2017). 

\bibitem{Yoneda}
Y.\,Yoneda, Phys. Rev. 131, 2010 (1963).

\bibitem{Sinha}
S.\,K.\,Sinha, E.\,B.\,Sirota, S.\,Garoff, and H.\,B.\,Stanley, Phys. Rev. B 38, 2297 (1988).

\bibitem{Vainshtein}
B.\,K.\,Vainshtein, {\it Diffraction of X-rays by Chain Molecules}, Elsevier, Amsterdam, 1966.

\bibitem{Vignaud}
G.\,Vignaud, A.\,Gibaud, J.\,Wang , S.\,K.\,Sinha , J.\,Daillant, G.\,Grubel, Y.\,Gallot, J. Phys.: Condens. Matter 9, L125 (1997).

\bibitem{MingLi}
M.\,Li, D.\,J.\,Chaiko, A.\,M.\,Tikhonov, M.\,L.\,Schlossman, Phys. Rev. Lett. 86, 5934 (2001).

\bibitem{MingLi2}
M.\,Li, A.\,M.\,Tikhonov, M.\,L.\,Schlossman, J. Europhys. Lett. 58, 80 (2002).

\bibitem{CW}
F.\,P.\,Buff, R.\,A.\,Lovett, F.\,H.\,Stillinger, Phys. Rev. Lett. 15, 621 (1965).

\bibitem{Schwartz}
D.\,K.\,Schwartz, M.\,L.\,Schlossman, E.\,H.\,Kawamoto, G.\,J.\,Kellogg, P.\,S.\,Pershan, B.\,M.\,Ocko,
Phys. Rev. A 41, 5687 (1990).

\bibitem{McClain}
B.\,R.\,McClain, D.\,D.\,Lee, B.\,L.\,Carvalho, S.\,G.\,J.\,Mochrie, S.\,H.\,Chen, J.\,D.\,Litster, Phys. Rev. Lett. 72, 246 (1994).

\bibitem{MWS}
D.\,M.\,Mitrinovic, S.\,M.\,Williams, M.\,L.\,Schlossman, Phys. Rev. E 63, 021601 (2001).

\bibitem{Weeks}
J.\,D.\,Weeks, J. Chem. Phys. 67, 3106 (1977).

\bibitem{Braslau2}
A.\,Braslau, M.\,Deutsch , P.\,S.\,Pershan, A.\,H.\,Weiss, J.\,Als-Nielsen, J.\,Bohr, Phys. Rev. Lett. 54, 114 (1985).

\bibitem{Tolan}
M.\,Tolan, {\it X-ray Scattering from Soft-Matter Thin Films}, Springer Tracts in Modern Physics 148, Springer, 1999.

\bibitem{22}
A.\,M.\,Tikhonov, J. Phys. Chem. B 110, 2746 (2006).

\bibitem{Tikh111}
A.\,M.\,Tikhonov, J. Phys. Chem. C 111, 930 (2007).

\bibitem{4}
D.\,M.\,Mitrinovic, A.\,M.\,Tikhonov, M.\,Li, Z.\,Huang, and M.\,L.\,Schlossman, Phys. Rev. Lett. 85, 582 (2000).

\bibitem{Small}
D.\,M.\,Small, {\it The Physical Chemistry of Lipids}, Plenum Press, New York, 1986.

\bibitem{Marchenko1}
V.\,I.\,Marchenko, JETP Lett. 33, 381 (1981).

\bibitem{Marchenko2}
V.\,I.\,Marchenko, Sov. Phys. JETP 54, 605 (1981).

\bibitem{CC}
A.\,M.\,Tikhonov, JETP 110, 1055 (2010). 

\bibitem{McConnell}
H.\,M.\,McConnell, Annu. Rev. Phys. Chem. 42, 171 (1991).

\bibitem{Uredat}
S.\,Uredat and G.\,Findenegg, Langmuir 15, 1108 (1999).

\bibitem{shuman}
A.\,W.\,Schuman, T.\,S.\,Bsaibes, M.\,L.\,Schlossman, Soft Matter 10, 7353 (2014).

\bibitem{MingLi3}
A.\,M.\,Tikhonov, M.\,Li, M.\,L.\,Schlossman, J. Phys. Chem. B 105, 8065 (2001).




\end{thebibliography}
\end{document}